\documentclass[final,5p,times,twocolumn]{elsarticle}

\usepackage{amsmath}
\usepackage{amssymb}
\usepackage{graphicx}
\usepackage{color}
\usepackage{hyperref}
\usepackage{xspace}

\usepackage[mathscr,scaled=1.15]{urwchancal}
\DeclareFontFamily{OT1}{pzc}{}
\DeclareFontShape{OT1}{pzc}{m}{it}%
{<-> s * [1.15] pzcmi7t}{}
\DeclareMathAlphabet{\mathpzc}{OT1}{pzc}{m}{it}

\definecolor{purple}{rgb}{0.5,0,0.5}
\definecolor{blue}{rgb}{0.0,0,0.9}

\biboptions{sort&compress}

\journal{Physics Letters B}

\begin{document}

\begin{frontmatter}

\title{Bridging a gap between continuum-QCD and \emph{ab initio} predictions of hadron observables}

\author[ECT]{Daniele Binosi}
\author[UA]{Lei Chang}
\author[UV]{Joannis Papavassiliou}
\author[ANL]{Craig D.~Roberts}

\address[ECT]{European Centre for Theoretical Studies in Nuclear Physics
and Related Areas (ECT$^\ast$) and Fondazione Bruno Kessler\\ Villa Tambosi, Strada delle Tabarelle 286, I-38123 Villazzano (TN) Italy}
\address[UA]{CSSM, School of Chemistry and Physics
University of Adelaide, Adelaide SA 5005, Australia}
\address[UV]{Department of Theoretical Physics and IFIC, University of Valencia and CSIC, E-46100, Valencia, Spain}
\address[ANL]{Physics Division, Argonne National Laboratory, Argonne, Illinois 60439, USA}

\date{11 December 2014}

\begin{abstract}
$\,$\\[-7ex]\hspace*{\fill}{\emph{Preprint no}.~ADP-14-42/T901}\\[1ex]
Within contemporary hadron physics there are two common methods for determining the momentum-dependence of the interaction between quarks: the top-down approach, which works toward an \emph{ab initio} computation of the interaction via direct analysis of the gauge-sector gap equations; and the bottom-up scheme, which aims to infer the interaction by fitting data within a well-defined truncation of those equations in the matter sector that are relevant to bound-state properties.  We unite these two approaches by demonstrating that the renormalisation-group-invariant running-interaction predicted by contemporary analyses of QCD's gauge sector coincides with that required in order to describe ground-state hadron observables using a nonperturbative truncation of QCD's Dyson-Schwinger equations in the matter sector.  This bridges a gap that had lain between nonperturbative continuum-QCD and the \emph{ab initio} prediction of bound-state properties.
%
\end{abstract}

\begin{keyword}
Dyson-Schwinger equations \sep
confinement \sep
dynamical chiral symmetry breaking \sep
fragmentation \sep
Gribov copies
\smallskip

\end{keyword}
\end{frontmatter}


\noindent\textbf{1.$\;$Introduction}.
%
The last two decades have seen significant progress and phenomenological success in the formulation and use of symmetry preserving methods in continuum-QCD for the computation of observable properties of hadrons \cite{Roberts:1994dr,Roberts:2000aa,Maris:2003vk,Eichmann:2009zx,Chang:2011vu,
Bashir:2012fs,Eichmann:2012zz,Cloet:2013jya}.  A large part of that work is based on the rainbow-ladder (RL) truncation of QCD's Dyson-Schwinger equations (DSEs), which is the leading-order term in a symmetry preserving approximation scheme \cite{Munczek:1994zz,Bender:1996bb}.  The RL truncation is usually employed with a one-parameter model for the infrared behaviour of the quark-quark interaction produced by QCD's gauge-sector \cite{Qin:2011ddS,Qin:2011xqS}.  It is accurate for ground-state vector- and isospin-nonzero pseudoscalar-mesons constituted from light quarks and also for nucleon and $\Delta$ properties because corrections in all these channels largely cancel owing to parameter-free preservation of the Ward-Green-Takahashi (WGT) identities \cite{Ward:1950xp,Green:1953te,Takahashi:1957xn,Takahashi:1985yz}.  Corrections do not cancel in other channels, however; and hence studies based on the RL truncation, or low-order improvements thereof \cite{Watson:2004kd,Fischer:2009jm}, have usually provided poor results for all other systems.

A recently developed truncation scheme \cite{Chang:2009zb} overcomes the weaknesses of RL truncation in all channels considered thus far.  This new strategy, too, is symmetry preserving but it has an additional strength; namely, the capacity to express dynamical chiral symmetry breaking (DCSB) nonperturbatively in the integral equations connected with bound-states.  That is a crucial advance because, like confinement, DCSB is one of the most important emergent phenomena within the Standard Model: it may be considered as the origin of more than 98\% of the visible mass in the Universe.  Owing to this feature, the new scheme is described as the DB truncation.  It preserves successes of the RL truncation but has also enabled a range of novel nonperturbative features of QCD to be demonstrated \cite{Chang:2010hb,Chang:2011ei,Chen:2012qrS,Chang:2013pqS}.

The widespread phenomenological success of this bottom-up approach to the calculation of hadron observables raises an important question; viz., are the one-parameter RL or DB interaction models, used in those equations relevant to colour-singlet bound-states, consistent with modern analyses of QCD's gauge sector and the solutions of the gluon and ghost gap equations they yield \cite{Aguilar:2008xm,Aguilar:2009nf,Aguilar:2009pp,Binosi:2009qm,Boucaud:2011ugS,
Pennington:2011xs,Maas:2011se,Cucchieri:2011ig,Dudal:2012zx,Strauss:2012dg,Aguilar:2013xqa}?  An answer in the affirmative will grant significant additional credibility to the claim that these predictions are firmly grounded in QCD.

\smallskip

\noindent\textbf{2.$\;$Quark gap equation}.
In order to expose the computational essence of the bottom-up DSE studies, it is sufficient to consider the gap equation for the dressed quark Schwinger function, $S(p)=Z(p^2)/[i\gamma\cdot p + M(p^2)]$:
\begin{subequations}
\label{gendseN}
\begin{align}
S^{-1}(p) & = Z_2 \,(i\gamma\cdot p + m^{\rm bm}) + \Sigma(p)\,,\\
\Sigma(p)& =  Z_1 \int^\Lambda_{dq}\!\! g^2 D_{\mu\nu}(p-q)\frac{\lambda^a}{2}\gamma_\mu S(q) \frac{\lambda^a}{2}\Gamma_\nu(q,p) ,
\end{align}
\end{subequations}
where: $D_{\mu\nu}$ is the gluon propagator;\footnote{
Landau gauge is typically used because it is, \emph{inter alia} \protect\cite{Bashir:2008fk,Bashir:2009fv,Raya:2013inaS}: a fixed point of the renormalisation group; that gauge for which sensitivity to model-dependent differences between \emph{Ans\"atze} for the fermion--gauge-boson vertex are least noticeable; and a covariant gauge, which is readily implemented in numerical simulations of lattice regularised QCD. Importantly, capitalisation on the gauge covariance of Schwinger functions obviates any question about the gauge dependence of gauge invariant quantities.}
$\Gamma_\nu$, the quark-gluon vertex; $\int^\Lambda_{dq}$, a symbol representing a Poincar\'e invariant regularisation of the four-dimensional integral, with $\Lambda$ the regularisation mass-scale; $m^{\rm bm}(\Lambda)$, the current-quark bare mass; and $Z_{1,2}(\zeta^2,\Lambda^2)$, respectively, the vertex and quark wave-function renormalisation constants, with $\zeta$ the renormalisation point, which is $\zeta_2=2\,$GeV here.  Eqs.\,\eqref{gendseN} are the starting point for all DSE predictions of hadron properties.

Significantly, owing to asymptotic freedom, there is no model dependence in the behaviour of the gap equation's kernel on the domain ${\mathpzc A} = \{(p^2,q^2)\,|\, k^2=(p-q)^2 \simeq p^2 \simeq q^2 \gtrsim 2\,{\rm GeV}^2\}$ because perturbation theory and the renormalisation group can be used to show \cite{Jain:1993qh,Maris:1997tm,Bloch:2002eq}:
\begin{align}
g^2 D_{\mu\nu}(k) \, Z_1 \, \Gamma_\nu(q,p)
& \stackrel{k^2\gtrsim 2\,{\rm GeV}^2}{=} 4\pi\alpha_s(k^2)\,
D^{\rm free}_{\mu\nu}(k) \, Z_2^2 \,\gamma_\nu\,,
\label{UVmodelindependent}
\end{align}
where $D^{\rm free}_{\mu\nu}(k)$ is the free-gauge-boson propagator and $\alpha_s(k^2)$ is QCD's running coupling on this domain.  Kindred results follow immediately for the kernels in the two-body Bethe-Salpeter equations relevant for meson bound-states \cite{Munczek:1994zz,Bender:1996bb,Chang:2009zb}.

Equation\,\eqref{UVmodelindependent} entails that the model input in realistic DSE studies is expressed in a statement about the nature of the gap equation's kernel on ${\mathpzc A}$; i.e., at infrared momenta.  One writes
\begin{subequations}
\label{kernelgap}
\begin{align}
& Z_1 g^2 D_{\mu\nu}(k) \Gamma_\nu(q,p) = k^2 {\cal G}(k^2)
D^{\rm free}_{\mu\nu}(k) \, Z_2\,\Gamma_\nu^{A}(q,p)\\
&=  \left[ k^2 {\cal G}_{\rm IR}(k^2) + 4\pi \tilde\alpha_{\rm pQCD}(k^2) \right]
D^{\rm free}_{\mu\nu}(k)\, Z_2\,\Gamma_\nu^A(q,p) ,
\label{realistic}
\end{align}
\end{subequations}
where $\tilde\alpha_{\rm pQCD}(k^2)$ is a bounded, monotonically-decreasing regular continuation of the perturbative-QCD running coupling to all values of spacelike-$k^2$; ${\cal G}_{\rm IR}(k^2)$ is an assumed form for the interaction at infrared momenta, with ${\cal G}_{\rm IR}(k^2)\ll \tilde\alpha_{\rm pQCD}(k^2)$ $\forall k^2\gtrsim 2\,$GeV$^2$; and $\Gamma_\nu^A(q,p)$ is an \emph{Ansatz} for the dressed-gluon-quark vertex, with $\Gamma_\nu^A(q,p) = Z_2 \gamma_\nu$ on ${\mathpzc A}$.

As reviewed elsewhere \cite{Chang:2011vu,Bashir:2012fs,Cloet:2013jya}, successful explanations and predictions of numerous hadron observables are obtained with
\begin{subequations}
\label{CalIGQC}
\begin{align}
\label{CalIQC}
{\mathcal I}(k^2) & = k^2 {\cal G}(k^2)\,,\\
\label{CalGQC}
{\cal G}(k^2) & = \frac{8 \pi^2}{\omega^4} D \, {\rm e}^{-k^2/\omega^2}
+ \frac{8 \pi^2 \gamma_m\, {\mathpzc E}(k^2)}{\ln [ \tau + (1+k^2/\Lambda_{\rm QCD}^2)^2]} ,
\end{align}
\end{subequations}
where: $\gamma_m = 12/(33-2 N_f)$ [typically, $N_f=4$], $\Lambda_{\rm QCD}=0.234\,$GeV; $\tau={\rm e}^2-1$; and ${\mathpzc E}(k^2) = [1 - \exp(-k^2/[4 m_t^2])]/k^2$, $m_t=0.5\,$GeV.  The origin and features of Eq.\,\eqref{CalGQC} are detailed in Ref.\,\cite{Qin:2011ddS} so here we only highlight two key aspects: the \emph{Ansatz} is consistent with the constraints described above and it involves just one free parameter.

The last point deserves further attention.  At first glance there appear to be two free parameters in Eq.\,\eqref{CalGQC}: $D$, $\omega$.  However, computations show  \cite{Maris:2002mt,Qin:2011ddS,Qin:2011xqS} that a large body of observable properties of ground-state vector- and isospin-nonzero pseudoscalar-mesons are practically insensitive to variations of $\omega \in [0.4,0.6]\,$GeV, so long as
\begin{equation}
 (\varsigma_G)^3 := D\omega = {\rm constant}.
\label{Dwconstant}
\end{equation}
(The midpoint $\omega=0.5\,$GeV is usually employed in calculations.)  This feature also extends to numerous properties of the nucleon and $\Delta$ resonance \cite{Eichmann:2009zx,Eichmann:2012zz}.  The value of $\varsigma_G$ is typically chosen in order to obtain the measured value of the pion's leptonic decay constant, $f_\pi$. It is striking that fitting just one parameter in a \emph{Gau{\ss}ian} \emph{Ansatz} for the gap equation's kernel is sufficient to achieve an efficacious description of a wide range of hadron observables.  It provides \emph{prima} \emph{facie} evidence that Eqs.\,\eqref{kernelgap}, \eqref{CalIGQC} are correct in principle; and translates the question posed at the end of Sect.\,1 into the following: ``How does $k^2 {\cal G}_{\rm IR}(k^2)$ in Eq.\,\eqref{CalIQC} compare with today's understanding of QCD's gauge sector?''

That question has a subtext, however, because the fitted value of $\varsigma_G$ depends on the form of $\Gamma_\nu^A(q,p)$.  We consider two choices herein: RL and DB.  The RL truncation is obtained with
\begin{equation}
\label{RLvertex}
\Gamma_\nu^A(q,p) = Z_2 \gamma_\nu\,.
\end{equation}
It is summarised in App.\,A.1 of Ref.\,\cite{Chang:2012cc} and provides the most widely used DSE computational scheme in hadron physics.  In this case one has \cite{Chang:2013pqS}
\begin{equation}
\varsigma_G^{\rm RL} = 0.87\,{\rm GeV.}
\label{sigmaRL}
\end{equation}

The form of $\Gamma_\nu^{\rm DB}(q,p)$ is detailed in App.\,A.2 of Ref.\,\cite{Chang:2012cc}.  It is consistent with constraints imposed by both the longitudinal and transverse WGT identities \cite{Qin:2013mtaS}.  The DB kernel is connected with the most refined nonperturbative truncation that is currently available.  It is therefore expected to be the most realistic.  With this vertex, one has \cite{Chang:2013pqS}
\begin{equation}
\varsigma_G^{\rm DB} = 0.55\,{\rm GeV.}
\label{sigmaDB}
\end{equation}

Following upon this discussion, we arrive at a pair of simple questions.  Does an analysis of QCD's gauge sector produce a running interaction-strength that is generally consistent with the form in Eqs.\,\eqref{CalIGQC}; and, if so, does it more closely resemble the function obtained with $\varsigma_G$ in Eq.\,\eqref{sigmaRL} or \eqref{sigmaDB}?

\smallskip

\noindent\textbf{3.$\;$RGI interaction kernel}.
In order to expose the quantity with which Eq.\,\eqref{CalGQC} should be compared, we must provide some background.  The Landau-gauge dressed-gluon propagator has the simple form
\begin{equation}
D_{\mu\nu}(k) = D_{\mu\nu}^{\rm free}(k)\Delta(k^2)= \left[\delta_{\mu\nu}-\frac{k_\mu k_\nu}{k^2} \right] \frac{\Delta(k^2)}{k^2}
=: {\mathpzc T}_{\mu\nu}^k {\mathpzc D}(k^2)
\,;
\label{Gprop}
\end{equation}
and since we are interested in QCD's gauge sector, the dressed-ghost propagator will also be relevant:
\begin{equation}
{\mathpzc F}(k^2) = - \frac{F(k^2)}{k^2}\,.
\label{ghostF}
\end{equation}

As we now explain, the scalar function in Eq.\,\eqref{ghostF} is connected with the following gluon-ghost vacuum-polarisation:
\begin{subequations}
\label{Lambdamunu}
\begin{align}
\Lambda_{\mu\nu}(k) & = \delta_{\mu\nu} G(k^2) + (k_\mu k_\nu/k^2) L(k^2)\\
& =  N_c
\int^\Lambda_{dq}\! g^2 {\mathpzc F}(k^2) \, D_{\mu\rho}(q) \,H_{\rho\nu}(q,k)\,,
\end{align}
\end{subequations}
which arises in contemporary applications of the pinch-technique (PT) \cite{Cornwall:1981zr,Cornwall:1989gv} to QCD's gauge sector \cite{Binosi:2002ez,Binosi:2009qm}.  The kernel $H_{\rho\nu}(q,k)$ in Eq.\,\eqref{Lambdamunu} is defined via
\begin{equation}
k_\nu H_{\mu\nu}(q,k) = \Gamma^{\mathpzc F}_\mu(q,k)\,,
\end{equation}
where $\Gamma^{\mathpzc F}_\mu(q,k)$ is the dressed-ghost-gluon vertex: in the absence of dressing, $H_{\mu\nu}(q,k) \to H_{\mu\nu}^0(q,k) = \delta_{\mu\nu}$ and $\Gamma^{\mathpzc F}_\mu(q,k)\to \Gamma_\mu^{{\mathpzc F}0}(q,k)= k_\mu$.  In Landau gauge, one has \cite{Aguilar:2009pp}
\begin{equation}
1/F(k^2) = 1 + G(k^2) + L(k^2).
\label{CBRST}
\end{equation}
Notably, Eq.\,\eqref{CBRST} is a consequence of QCD's BRST invariance; and the combination $[1+G(k^2)]$ is a crucial element in the set of background-quantum identities (BQIs) explained in Ref.\,\cite{Binosi:2002ez}.

The functions $G(k^2)$ and $L(k^2)$ both satisfy dynamical equations \cite{Binosi:2009qm}, which may readily be deduced from the definitions in Eqs.\,\eqref{Lambdamunu}.  Analysing this dynamical system, one can prove the exact result \cite{Aguilar:2009pp} $L(0) =0$; and hence, using Eq.\,\eqref{CBRST}: $[1+G(k^2=0)=1/F(0)]$.  However, $L(k^2)\not\equiv 0$.  In fact, as demonstrated elsewhere \cite{Aguilar:2013xqa}, $L(k^2)$ is sizeable at intermediate momenta, so that the oft used assumption $[1+G(k^2)\approx 1/F(k^2)]$ is quantitatively inaccurate.  Therefore, in computing $G(k^2)$ herein, we include the nonperturbative corrections to $\Gamma^{\mathpzc F}_\mu(q,k)$ determined in Ref.\,\cite{Aguilar:2013xqa}.

As remarked above and elucidated elsewhere \cite{Munczek:1994zz,Bender:1996bb,Chang:2009zb}, there is a one-to-one correspondence between the kernels in the dressed-quark gap equation and those in the Bethe-Salpeter equations relevant to meson bound-states, which we will denote by ${\mathpzc K}$.  Crucially, the kernels ${\mathpzc K}$ possess a ``universal'' subcomponent, which has the nature of a running interaction-strength (coupling),  $\hat d(k^2)$, that does not depend on the valence-quark content of the Bethe-Salpeter equation.

A systematic identification of $\hat d(k^2)$ has been completed using the pinch technique \cite{Aguilar:2009nf}.  It was achieved via the rearrangement of physical amplitudes into sub-amplitudes with special properties.  In this way one obtains dressed coloured vertices that satisfy QED-like WGT identities and a gluon propagator that captures all the theory's renormalisation-group logarithms.  These quantities coincide with the corresponding vertices and propagator defined in the background field method (BFM) \cite{Abbott:1980hw}.  This identification is valid both perturbatively, to all orders, and nonperturbatively, at the level of the corresponding DSEs.  Moreover, the relationship between the corresponding Schwinger functions before and after the diagrammatic application of the PT procedure is formally captured by the BQIs.

To be specific, in the present context the standard gluon dressing function, $\Delta(k^2)$ in Eq.\,\eqref{Gprop}, is related to the scalar cofactor of the PT-BFM gluon propagator, denoted $\hat\Delta (k^2)$, as follows:
\begin{equation}
\Delta(k^2) = \hat\Delta(k^2) [1+G(k^2)]^2,
\label{BQI}
\end{equation}
where $G(k^2)$ was introduced in Eqs.\,\eqref{Lambdamunu}.  Evidently, $\hat\Delta(k^2)$ is related to $\Delta(k^2)$ via a function determined by ghost-gluon dynamics.

Similarly, the PT-BFM quark-gluon vertex: $\hat\Gamma^a_\mu = \frac{\lambda^a}{2} \hat\Gamma_\mu$, which satisfies a QED-like WGT identity:
\begin{equation}
k_\mu i \hat{\Gamma}_{\mu}(p,q)= S^{-1}(p) - S^{-1}(q),
\label{WI}
\end{equation}
is related to $\Gamma_\mu$ in Eqs.\,\eqref{gendseN} by the BQI
\begin{align}
[1+G(k^2)]& \Gamma_\mu(p,q) = \hat{\Gamma}_\mu(p,q) \nonumber\\
& + S^{-1}(q) Q_{\mu}(p,q)+\bar Q_{\mu}(p,q)S^{-1}(p),
\label{BQI-final}
\end{align}
where $Q_{\mu}$, $\bar Q_{\mu}$ are auxiliary composite three-point functions.  Importantly, when embedded in the computation of scattering processes, the last two terms on the right-hand-side of Eq.\,\eqref{BQI-final} cancel against other process-dependent contributions.

These considerations entail that the leading term in the quark-antiquark scattering kernel ($q_+ - q_- = P = p_+-p_-$)
\begin{align}
{\mathpzc K}(p,q;P)
=
    -\frac{\lambda^a}{2}\Gamma_\mu(p_+,q_{+}) \,
    \alpha_s D_{\mu\nu}(k)\,
    \frac{\lambda^a}{2}\Gamma_\nu(q_{-},p_-),
\label{fullK}
\end{align}
in which the spinor indices have been suppressed, may be rewritten as
\begin{align}
{\mathpzc K}(p,q;P)
=
    -\frac{\lambda^a}{2}\hat \Gamma_\mu(p_+,q_{+})\,
    \alpha_s \hat\Delta(k^2) D_{\mu\nu}^{\rm free}(k) \,
    \frac{\lambda^a}{2}\hat \Gamma_\nu(q_{-},p_-)\,.
\label{fullKhat}
\end{align}
Given that $\hat \Gamma_\nu$ in Eq.\,\eqref{fullKhat} satisfies Eq.\,\eqref{WI}, which is a significant element in the construction of the RL and DB truncations described above, then a comparison with Eqs.\,\eqref{kernelgap} and their computable analogues for the Bethe-Salpeter equations \cite{Munczek:1994zz,Bender:1996bb,Chang:2009zb} leads one to conclude that the dimensionless quantity
\begin{equation}
{\mathcal I}_{\hat d}(k^2) := k^2 \hat d(k^2) = \frac{\alpha_s(\zeta^2)\Delta(k^2;\zeta^2)}{[1+G^2(k^2;\zeta^2)]^2}
\label{eqdk2}
\end{equation}
is the object that should be compared directly with the interaction in Eq.\,\eqref{CalIQC}.

\begin{figure}[t]

\centerline{\includegraphics[width=0.8\linewidth]{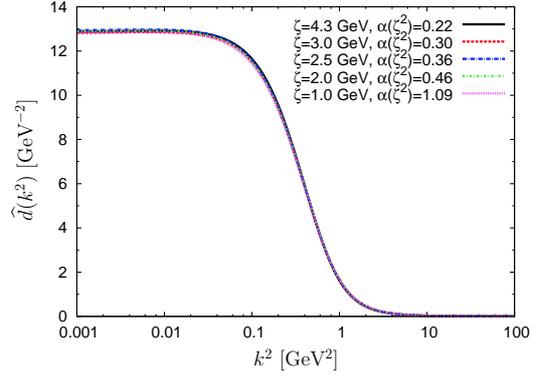}}

\caption{RGI running interaction strength, $\hat d(k^2)$ in Eq.\,\eqref{eqdk2}, computed via a combination of DSE- and lattice-QCD results, as explained in  Ref.\,\cite{Aguilar:2009nf}.  We display the function obtained using five different values of the renormalisation point in order to highlight that the result is RGI.
\label{Fighatd}}
\end{figure}

It is noteworthy that $\hat d(k^2)$ is a renormalisation-group-invariant (RGI) \cite{Aguilar:2009nf}.  The interaction defining the DB kernel, which is built using a sophisticated dressed-quark-gluon vertex, shares this feature.  In contrast, the RL kernel and pointwise improvements thereof, possesses some residual $\zeta$-dependence because the structure of $\Gamma_\mu^A$ is too simple.

\smallskip

\noindent\textbf{4.$\;$Numerical results for the RGI kernel}.
The best available information on the RGI running interaction-strength in Eq.\,\eqref{eqdk2} has been obtained through a combination of DSE- and lattice-QCD analyses.  The procedure is detailed in Ref.\,\cite{Aguilar:2009nf} and yields the  result depicted in Fig.\,\ref{Fighatd}.  We do not review the method herein; but, since the renormalisation-point-independence of the result is important, we recapitulate upon relevant aspects of that procedure, which are elaborated elsewhere \cite{Aguilar:2009pp}.  One begins with lattice-QCD results for the ghost dressing function \cite{Bogolubsky:2009dc}, ${\mathpzc F}(k^2;\zeta_{4.3}^2)$ in Eq.\,\eqref{ghostF}, which was renormalised to unity at $\zeta=4.3\,$GeV, and determines $\alpha_s(\zeta_{4.3}^2)$ by requiring that the DSE for ${\mathpzc F}(k^2;\zeta_{4.3}^2)$ reproduces the lattice result.  The value of $\alpha_s$ at a new value of $\zeta$ is then obtained by employing multiplicative renormalisability in order to rescale the lattice result and subsequently fixing $\alpha_s(\zeta^2)$ so that the gap equation for ${\mathpzc F}(k^2;\zeta^2)$ reproduces the rescaled lattice function.  In following this procedure, all scales are completely determined by the original lattice result and the running of $\alpha_s(\zeta^2)$ matches that of perturbative QCD on the perturbative domain $(\zeta^2\geq \zeta_2^2)$; viz., the four-loop expression for the running coupling evaluated in the momentum-subtraction scheme with a value of $\Lambda_{\rm QCD}$ between $0.25$ and $0.32\,$GeV \cite{Boucaud:2005gg,Boucaud:2008gn}.

One may naturally separate $\hat d(k^2)$ into the product of two scale-dependent terms: a dimensionless running coupling, $\alpha(\zeta^2)$, multiplied by the mass-dimension ``$-2$'' PT-BFM gluon propagator, $\hat \Delta(k^2)/k^2$, Eq.\,\eqref{BQI}.  The latter quantity can be used to define a $\zeta$-dependent gluon mass-scale:
\begin{equation}
m_g^2(\zeta^2) = \lim_{k^2\to 0} [1/{\mathpzc D}(k^2) = k^2/\hat \Delta(k^2;\zeta^2)].
\end{equation}
It is evident from Fig.\,\ref{Fighatd} that with $\alpha_s(\zeta)$ finite and nonzero, then $0<m_g^2(\zeta^2)<\infty$.

\begin{figure}[t]

\centerline{\includegraphics[width=0.85\linewidth]{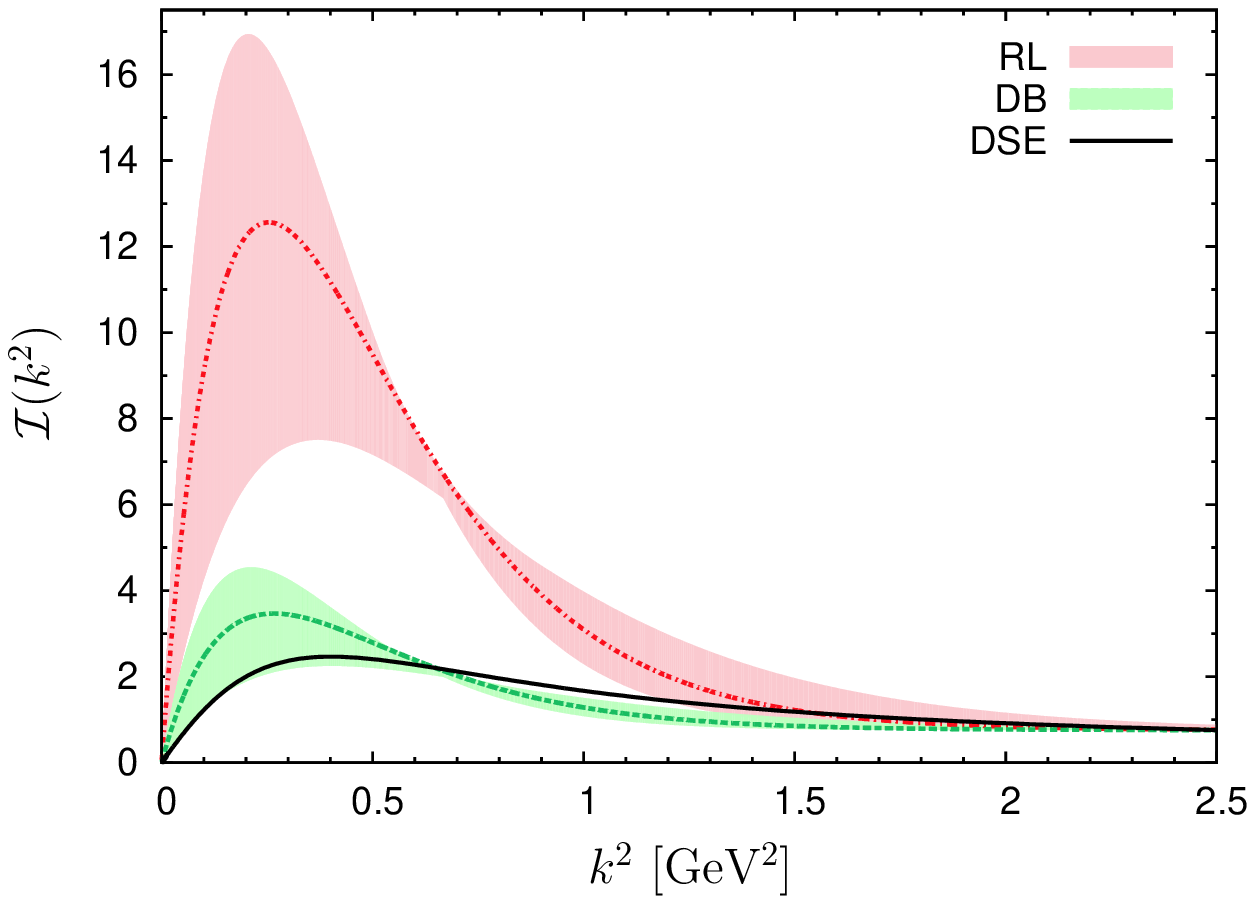}}
\centerline{\includegraphics[width=0.85\linewidth]{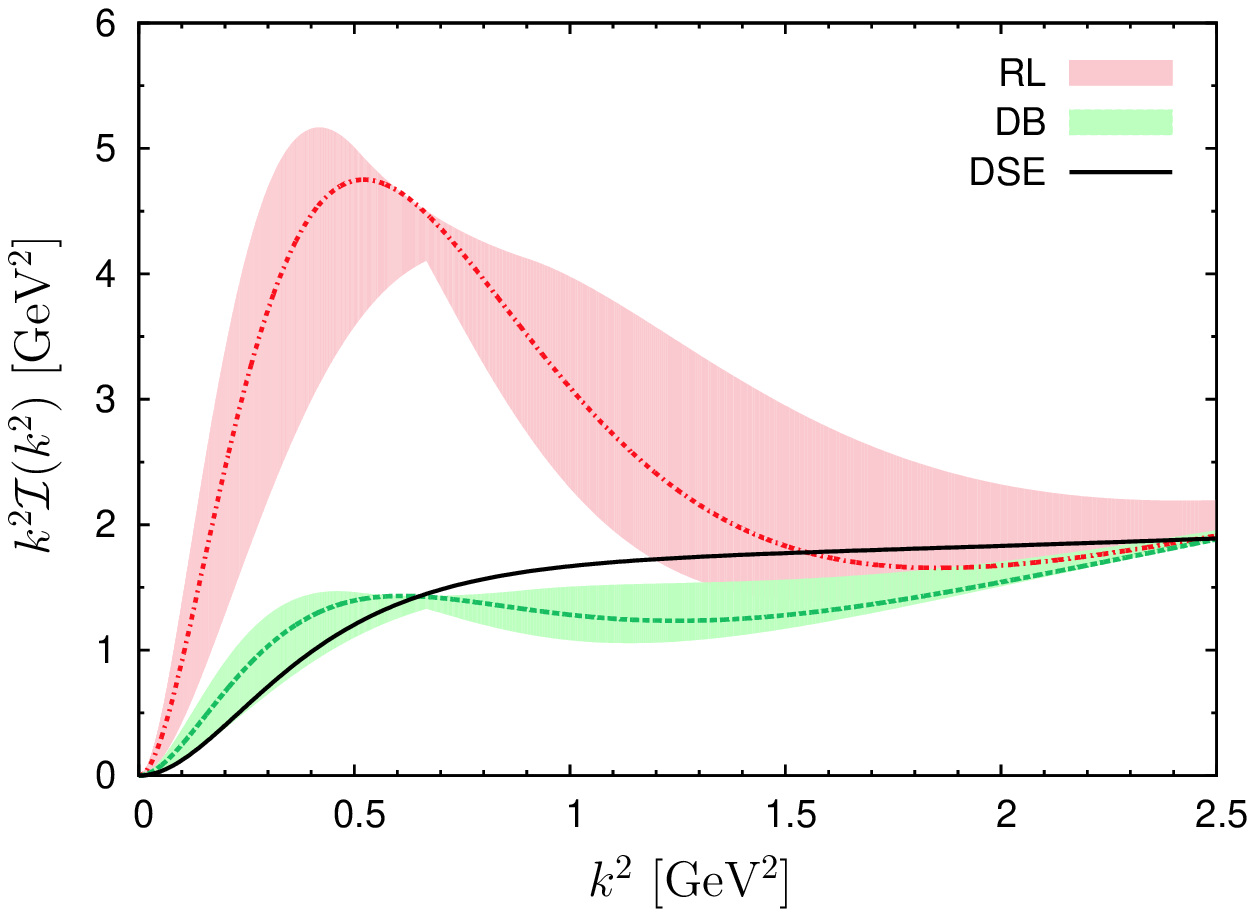}}

\caption{Comparison between top-down results for the gauge-sector interaction [Eqs.\,\eqref{eqdk2}, \eqref{gluonmass}, Fig.\,\ref{Fighatd}] with those obtained using the bottom-up approach based on hadron physics observables [Eqs.\,\eqref{CalIGQC}--\eqref{sigmaDB}].  \emph{Solid curve} -- top-down result for the RGI running interaction; \emph{pale-red band} -- bottom-up result obtained in the RL truncation, Eqs.\,\eqref{CalIGQC} \& \eqref{sigmaRL}; and \emph{pale-green band} -- bottom-up result obtained in the DB truncation, Eqs.\,\eqref{CalIGQC} \& \eqref{sigmaDB}.  The bands denote the domain of constant ground-state physics $0.45<\omega<0.6$, described in connection with Eq.\,\eqref{Dwconstant}, with $\omega=0.6$ producing the ``flattest'' curve; i.e., the smaller values on $k^2\lesssim 0.65\,$GeV$^2$ and the larger values on $k^2\gtrsim 0.65$GeV$^2$.  The curve within each band is obtained with $\omega=0.5\,$GeV.  All curves are identical on the perturbative domain: $k^2>2.5\,$GeV$^2$.
\label{TopBottom}}
\end{figure}

Repeating the analysis in Ref.\,\cite{Aguilar:2009nf}, we obtain $\hat d(k^2=0)=12.9/{\rm GeV}^2$, as is plain from Fig.\,\ref{Fighatd}; and our results on the domain $1<\zeta<4.3\,$GeV are accurately interpolated by
\begin{equation}
m_g^2(\zeta^2) = \frac{0.22 + 0.019 \,\zeta^2}{1+1.76 \,\zeta^2}\,,\;
\alpha_s(\zeta^2) = 12.9\, m_g^2(\zeta^2)\,.
\end{equation}
Extrapolating to the far infrared, one finds that QCD's gauge sector is characterised by the following coupling and mass-scale:
\begin{equation}
\alpha_s(0) = 2.77 \approx 0.9  \pi \,,\;
m_g^2(0) = (0.46\,{\rm GeV})^2\,.
\label{gluonmass}
\end{equation}
The value of the mass-scale in Eq.\,\eqref{gluonmass} is \emph{natural} in the sense that it is commensurate with but larger than the value of the dressed light-quark mass function at far-infrared Euclidean momenta: $M(0)\approx 0.3\,$GeV \cite{Bhagwat:2006tu}.  The strength of the coupling is also interesting.  It is greater than that required for DCSB to occur in simple treatments of strong-coupling QED \cite{Johnson:1964da,Fukuda:1976zb} ($\alpha_c=\pi/3$) and gap equation models for QCD \cite{Higashijima:1983gx,Roberts:1989mj} ($\alpha_c \approx \pi/3.5$), and consistent with the value often imagined necessary to describe strong-interaction phenomena: $\alpha_s(0)\gtrsim \pi$ (see, e.g., Refs.\,\cite{Brodsky:2010ur,Roberts:2011wyS,Qin:2011ddS}).

\smallskip

\noindent\textbf{5.$\;$Comparison of interaction kernels}.
It is now possible to compare the prediction yielded by analyses of QCD's gauge sector (top-down approach) with the running-interaction determined using the bottom-up approach; i.e., parametrising the gauge-sector kernel and fitting the parameter in order to explain a wide range of hadron observables.  The results of the comparison are displayed in Fig.\,\ref{TopBottom}: the upper panel depicts ${\mathcal I}(k^2)$ itself, whereas the lower panel portrays $k^2 {\mathcal I}(k^2)$.  The latter is plotted because the computation of hadron observables typically involves a four-dimensional Euclidean integration, which introduces an additional factor of $k^2$ from the measure.

It is immediately apparent that the top-down RGI interaction (solid-black curve) and the DB-truncation bottom-up interaction (green band containing dashed curve) are in excellent agreement.  Hence, the interaction predicted by modern analyses of QCD's gauge sector is in near precise agreement with that required for a veracious description of measurable hadron properties using the most sophisticated matter-sector gap and Bethe-Salpeter kernels available today. This is a remarkable result given that, on the themes described herein, there had previously been no communication between the continuum-QCD hemisphere represented by the studies described in Sect.\,2 and that connected with Sects.\,3 and 4, except insofar as it was mutually acknowledged that QCD's gauge sector is characterised by a nonzero and finite gluon mass-scale.

Unlike one of its predecessors \cite{Maris:1999nt}, the modern interaction inferred using RL-truncation \cite{Qin:2011ddS,Qin:2011xqS} has the correct shape; but it is too large in the infrared.  This is because the bare-vertex in Eq.\,\eqref{RLvertex} suppresses all effects associated with DCSB in the kernels of the gap and Bethe-Salpeter equations \emph{except} those expressed in ${\mathcal I}(k^2)$, and therefore a description of hadronic phenomena can only be achieved by overmagnifying the gauge-sector interaction strength at infrared momenta.  A similar conclusion was drawn elsewhere \cite{Rojas:2014tya}.  It follows that whilst the RL truncation supplies a useful computational link between QCD's gauge sector and measurable hadron properties, the model interaction it delivers should neither be misconstrued nor misrepresented as a pointwise-accurate representation of ghost-gluon dynamics.  Notwithstanding this, the judicious use of RL truncation and the careful interpretation of its results can still be a valuable tool for hadron physics phenomenology.\footnote{The shortcomings of the interaction in Ref.\,\cite{Maris:1999nt} are explained in Ref.\,\cite{Qin:2011ddS}.  They are practically immaterial if one only considers low momentum transfer properties of hadron ground-states, which are mainly sensitive to the integrated strength of the interaction on the infrared domain: $k^2\lesssim 2\,$GeV$^2$.  This fact is highlighted by the phenomenological successes of a RL treatment of a vector$\,\otimes\,$vector contact-interaction (see Refs.\,\cite{Roberts:2011wyS,Chen:2012qrS,Pitschmann:2014jxa} and citations therein).
}

\begin{table}[t]
\caption{Row~1 -- Computed values determined from the interaction tension in Eq.\,\eqref{InteractionTension}, quoted in GeV; and Row~2 -- the difference: $\varepsilon_\varsigma:= \surd \varsigma^{\mathcal I}/\surd \varsigma^{{\mathcal I}_{\hat d}}-1$.  So as to represent the domain of constant ground-state physics, described in connection with Eq.\,\eqref{Dwconstant}, we list values obtained with bottom-up interactions using $\omega=0.5$, $0.6\,$GeV.
\label{interactiontension}}
\begin{tabular*}
{\hsize}
{
l|
l@{\extracolsep{0ptplus1fil}}
l@{\extracolsep{0ptplus1fil}}
l@{\extracolsep{0ptplus1fil}}
l@{\extracolsep{0ptplus1fil}}
l@{\extracolsep{0ptplus1fil}}}\hline
${\mathcal I}$ & ${\mathcal I}_{\hat d}$ &
${\mathcal I}_{\rm DB}^{\omega=0.5}$ &
${\mathcal I}_{\rm DB}^{\omega=0.6}$ &
${\mathcal I}_{\rm RL}^{\omega=0.5}$ &
${\mathcal I}_{\rm RL}^{\omega=0.6}$ \\\hline

$\surd\varsigma^{\mathcal I}$
& $1.86$ & $1.91$ & 1.82 & $3.14$ & 2.90 \\
$\varepsilon_\varsigma$& 0 & $2.8$\% & \rule{-1.3ex}{0ex}$-2.4$\% & 68.5\% & 55.8\%\\\hline
\end{tabular*}
\end{table}

The level of agreement between the curves in Fig.\,\ref{TopBottom} can usefully be quantified via the interaction tension \cite{Roberts:2000hi}
\begin{equation}
\label{InteractionTension}
\sigma^{\mathcal I} := \int_0^{k_p^2} \! dk^2 \, {\mathcal I}(k^2)\,,
\end{equation}
where the integral is limited to $k^2\leq k_p^2=2.5\,$GeV$^2$ because all interactions are identical on the perturbative domain.  The results are listed in Table~\ref{interactiontension}.  According to the metric defined by the second row, there is plainly little measurable difference between the top-down prediction and the DB-truncation bottom-up interaction.  This is true independent of whether one uses ${\mathcal I}(k^2)$ or $k^2{\mathcal I}(k^2)$ as the integrand.

\smallskip

\noindent\textbf{6.$\;$Confinement, fragmentation and Gribov copies}.
It has long been known that the only Schwinger functions which can be associated with states in the Hilbert space of observables; namely, the set of measurable expectation values, are those that satisfy the axiom of reflection positivity \cite{Krein:1990sf}.  In this connection, it is a simple matter to show that the spectral density associated with any single-variable Schwinger function, ${\mathpzc S}(k^2)$, which possesses an inflexion point at $k^2>0$ cannot be positive definite \cite{Aiso:1997au,Roberts:2007ji}; i.e., it violates the axiom of reflection positivity and hence the associated excitation may be viewed as \emph{confined}.

The RGI function $\hat d(k^2)$ displayed in Fig.\,\ref{Fighatd} possesses and inflection point at $k^2=k_i^2=(0.41\,{\rm GeV})^2$.  Consequently, when probed at momenta $k^2>k_i^2$, the PT-BFM gluon behaves as an ``ordinary'' excitation because on this domain its propagator has the convex shape that is characteristic of free-particle-like behaviour.  Notably, the \emph{computed} value $|k_i|=0.41\,{\rm GeV}$ corresponds to a length ${\mathpzc s}_i \approx 0.5\,{\rm fm}$, which is a natural scale for confinement in QCD; and as $k^2$ decreases through $k_i^2$, passing into the infrared, the effects of strong, nonperturbative ghost and gluon dressing become manifest and the excitation's propagation characteristics are dramatically altered.

Indeed, as described twenty years ago \cite{Stingl:1994nk}, a violation of positivity owing to the dynamical generation of a length-scale ${\mathpzc s}_i \approx 0.5\,{\rm fm}$ may plausibly be connected with the fragmentation phenomenon.  Namely, the coloured state propagates as a pseudo-plane-wave over mean-distances $\langle{\mathpzc s}\rangle <{\mathpzc s}_i$.  However, after each ``step'' of length ${\mathpzc s}_i$, on average, an interaction occurs, so that the coloured state loses its identity, sharing it with other partons.  Finally, a cloud of partons is produced, which coalesces into a number (often large) of colour-singlet final states.  This realisation of confinement is essentially dynamical.  It is not connected in any way with the static potential between infinitely-heavy quarks measured in numerical simulations of quenched lattice-QCD.\footnote{This static potential is irrelevant to the question of confinement in a universe in which light quarks are ubiquitous because light-particle creation and annihilation effects are essentially nonperturbative and therefore it is impossible in principle to compute a (non light-front) quantum mechanical potential between two light quarks \cite{Bali:2005fuS,Chang:2009aeS}.}

The fact that the active piece of the gauge-boson Schwinger function acquires a dynamically generated mass-scale ensures that gluons with wavelengths $\lambda \gg 1/m_g(0)\approx 0.5\,$fm play no role in hadron observables.  This phenomenon, and the analogue for quarks, provides a basis for understanding the notion of a maximum wavelength for gluons and quarks in QCD \cite{Brodsky:2008be}.

One of the strengths of the framework we have employed is the intimate connection it draws between confinement, DCSB, dynamically generated gluon and quark masses, and the maximum wavelengths for gluons and quarks.  We have exemplified this above and choose to highlight another example here.  Namely, the behaviour of $\hat d(k^2)$ entails DCSB, which itself guarantees that light-quark dynamics in QCD supports the existence of a (nearly-)massless pseudoscalar meson; viz., the pion, whose properties are almost entirely determined by the dressed-quark mass-function \cite{Maris:1997hd}.  The exceptionally light pion degree-of-freedom becomes dominant in QCD at those length-scales above which dressed-gluons and -quarks decouple from the theory owing to the large magnitudes of their dynamically generated masses.  We therefore judge that Gribov copies can have no measurable impact on observables within the Standard Model because they affect only those gluonic modes whose wavelengths lie in the far infrared; and such modes are dynamically screened, by an exponential damping factor $\sim \exp(-\lambda m_g)$, so that their role in hadron physics is superseded by the dynamics of light-hadrons.  This conjecture is consistent with the insensitivity to Gribov copies of the dressed-quark and -gluon two-point Schwinger functions observed in numerical simulations of QCD on fine lattices \cite{Bowman:2002fe,Zhang:2004gv}.

\smallskip

\noindent\textbf{7.$\;$Epilogue}.
We have demonstrated that the form of the renormalisation-group-invariant running-interaction predicted by contemporary analyses of QCD's gauge sector is a good match to the behaviour required in order to describe a wide range of hadron observables using the most sophisticated, nonperturbative truncation of QCD's Dyson-Schwinger equations in the matter sector that is currently available.  In doing so, we have drawn a direct connection between QCD's gauge sector and measurable hadron properties.  This paves the way for genuinely \emph{ab initio} predictions of observables in continuum-QCD.

The understanding highlighted by Fig.\,\ref{TopBottom} was only made possible by recent progress in developing nonperturbative truncations for the gap and bound-state equations in QCD's matter sector.  The new symmetry-preserving scheme enables the influence of dynamical chiral symmetry breaking (DCSB) and concurrent phenomena to be spread throughout the gap and Bethe-Salpeter kernels so that the universal subcomponent associated cleanly with the gauge-sector can reliably be separated.  The primary element in the new scheme is an accurate representation of the dressed-quark-gluon vertex, $\Gamma_\mu$.  Our analysis thus emphasises the need for a continuation and expansion of efforts to better determine $\Gamma_\mu$, both in continuum- and lattice-QCD.

\smallskip

\noindent\textbf{Acknowledgments}.
We thank
A.\,C.~Aguilar,
I.\,C.~Clo\"et,
B.~El-Bennich, 
M.\,R.~Pennington,
M.~Pitschmann,
J.~Rodr\'iguez-Quintero,
J.~Segovia,
P.\,C.~Tandy 
and
A.\,W.~Thomas
for valuable discussions and suggestions.
DB, JP and CDR are grateful for the chance to participate in the workshops ``DSEs in Modern Physics and Mathematics,'' ECT$^\ast$, Villazzano, Trento, Italy, and ``Connecting Nuclear Physics and Elementary Particle Interactions: Building Bridges at the Spanish Frontier,'' Punta Umbr\'ia, Spain, during which this work was conceived and begun.
This research was supported by:
University of Adelaide and Australian Research Council through grant no.~FL0992247;
Spanish MEYC grant no.\ FPA2011-23596;
Generalitat Valenciana grant ``PrometeoII/2014/066'';
and U.S.\ Department of Energy, Office of Science, Office of Nuclear Physics, contract no.~DE-AC02-06CH11357.




\end{document}